# ANALYTICAL PROPERTIES OF THE PHOTON PROPAGATOR AND CONFINEMENT IN QED$_3$.

## M.SH.PEVZNER

Department of Physics, National Mining University of Ukraine,

19, Karl Marx Avenu, Dniepropetrovsk, 49600, Ukraine

E-mail: PevznerM@nmuu.dp.ua, mark@omp.dp.ua

Abstract

Analytical properties of the photon propagator in QED$_3$ in the $N^{-1}$-approximation are considered. It is noticed that in given approximation if the dynamical fermions mass is nonzero this propagator has a single pole at the point $k^2=0$ and a branch point at $k^2=-4m^2$ ( $m$ is the fermion mass). When $m=0$, the propagator gets the pole at the spacelike domain of the variable $k^2$. It witnesses the unstability of the state conforming to the photon and leads to the screening of the source field at $r \to \infty$. The approximation by the rational function of the photon propagator containing the vacuum loops with massive fermions proposed by Gusynin, Hams and Reenders is shown to lead to the pole and the branch point at $k^2=0$ and to the pole in the spacelike domain as well. The arguments in favour of the statement that the presence of the dynamical fermions mass leads in QED$_3$ to the confinement irrespectively of any approximation, are given.

Confinement is one of the most interesting phenomena of the modern quantum field theory. There are various definitions of confinement. We shall proceed from one, which considers confinement as the growth of the static point charge potential modulus at $r \to \infty$. In case of the quantum electrodynamics in three-dimensional space (QED$_3$) it takes place already at the tree approximation, so that when one investigates confinement address to QED$_3$ is justified since in virtue of the model simplicity it is possible to advance further in the problem understanding. The various aspects of the problem were considered recently in $[1-4]$. This communication supplements and specifies the results contained in the works mentioned above.

1. a) Let us examine the case when the dynamical fermions are massless. Then, retaining the first nonvanishing term of the polarization operator $N^{-1}$ expansion ($N$ is the number of fermions), for the transverse photon propagator part regularized by the gauge invariant manner, we have [5]

$$D_R^{(t)}(k^2) = \frac{1}{i(k^2 + (\alpha/8)k)}; \qquad (1)$$

here $k = (k^2)^{1/2}$, $\alpha = e_0^2 N$, $e_0^2$ is the nonrenormalized coupling constant in QED$_3$.

It is possible to rewrite the expression (1) in the equivalent form

$$D_R^{(t)}(k^2) = \frac{1}{i} \frac{1}{k^2 - (\alpha/8)^2} - \frac{1}{i} \frac{(\alpha/8)(k^2)^{1/2}}{k^2 - (\alpha/8)^2}. \tag{2}$$

It is seen from (2) that the function $D_R^{(t)}(k^2)$ considered in the complex plane of the variable $z = k^2$ has the pole at $k^2 = (\alpha/8)^2$ and the branch point at $k^2 = 0$. The pole of given function is in the spacelike domain of the $k^2$ meanings, that signifies the unstability of the single-photon state. The stable single-photon state shows itself by the pole presence at $k^2 = 0$ (and as a consequence of this by the presence of the term $A_0(r) \sim \ln r$ in the expression for the point charge static potential signaling the confinement existence). Thus the screening of the source field at $r \to \infty$ is the consequence of the single-photon state unstability at the approximation considered for this model.

b) Now we shall consider the case of the massive dynamical fermions. In this case it is possible to write the propagator $D_R^{(t)}(k^2)$ calculated in [5] in the form

$$D_R^{(t)}(k^2) = \frac{1}{i} \frac{1}{k^2 F(k^2)}, \tag{3}$$

where $F(k^2) = 1 + \frac{\alpha m}{2k^2} \left( 1 + \frac{1}{2}\left(\frac{k^2}{4m^2} - 1\right)\left(-\frac{4m^2}{k^2}\right)^{1/2} \ln \frac{1 + (-k^2/4m^2)^{1/2}}{1 - (-k^2/4m^2)^{1/2}} \right)$. In the complex plane of the variable $z = k^2$ this function has a simple pole at the point $k^2 = 0$ with the residue $F^{-1}(0) = \frac{1}{1 + \alpha/6\pi m}$ and it has a branch at the point $k^2 = -4m^2$. Besides that it has some other singularities at the points defined by the relation $F(k^2) = 0$. These singularities are not considered here.

The presence of the pole of the function $D_R^{(t)}(k^2)$ at $k^2 = 0$ providing confinement is the most essential feature of this case. When $m \to 0$ the residue at the pole is turned into zero and we come to the result discussed above.

c) As the general form of the function $F(k^2)$ at the approximation under consideration it remains rather complicated and it is approximated by more simple functions for solving concrete problems. As an example we shall consider the approximation used in [3]. In this case it is possible to represent the function $F(k)$ in the form

$$F(k^2) = \frac{\sqrt{k^2} + \alpha/8 + 3\pi m/4}{\sqrt{k^2} + 3\pi m/4} \tag{4}$$

then for the photon propagator we have

$$D_R^{(t)}(k^2) = \frac{1}{iF(0)}\frac{1}{k^2} + \frac{\alpha}{i6\pi m}\frac{1}{F(0)}\frac{1}{k^2 - (3/4\pi m F(0))^2} - \frac{\alpha}{i8}\frac{(k^2)^{1/2}}{k^2 - (3/4\pi m F(0))^2} . \tag{5}$$

Here we have the photon pole as well as the unphysical pole at the point $k^2 = (\alpha/8)^2$, related to the unstability of the single-photon state. The symmetry dynamical breaking takes place if $\alpha > m$. Nevertheless the photon pole suppresses the cjntribution of the other terms at $r \to \infty$ and the confinement takes place.

    2. From the simple intuitive considerations it is possible to state that the influence of the dynamical fermions mass on the static source field behaviour at $r \to \infty$ is not an approximation artefact but it has deep physical nature. Really, already from the classical dispersion formula for dielectrical permeability of the medium it follows that at $m \to 0$ the electrical field in the dielectric is completely screened. In $QED_4$ for the renormalized charge the following relation takes place

$$e^2 = \frac{e_0^2}{1 + (e_0^2/12\pi^2)\ln(L^2/m^2)} \tag{6}$$

($L$ is the cutoff momentum). From the relation (6) it follows that if $m = 0$ the "bare" charge is also completely screened at any distance from the source.

    It is possible to advance some the general reasons, not depending on any approximation that the nonzero dynamical fermion mass leads to confinement. Let us make the following assumptions: (i) there are asymptotical states of Hamiltonian in $QED_3$ corresponding to a photon and an arbitrary $e^+e^-$ pair aggregation; (ii) the state nearest to the single-photon one is separated from it by a $2m$ wide gap; (iii) the charge renormalization constant $Z_3 \neq 0$. These assumptions exclude the confinement since one of its definitions implies the absence of the particles of asymptotical states corresponding to the mass-shell (the analytical properties of the propagators in the

gauges theories with confinement were considered in [6] ). Then for the photon propagator the Källen – Lehmann representation will be just in the form [7,8]

$$D_R^{(t)}(k^2) = \frac{z}{i}\left(\frac{1}{k^2 - i0} + \int_{4m^2}^{\infty}\frac{\rho(\kappa^2)}{k^2 + \kappa^2 - i0}d\kappa^2\right). \tag{7}$$

Since $\rho(\kappa^2) \geq 0$ and $Z_3 \neq 0$ ( the assumption (iii)) the photon pole does not disappear when one performs the integration what leads to the confinement. Thus our assumption (ii) contradicts with the obtained result that is an additional argument in behalf of the confinement existence in $QED_3$ when the dynamical fermions have the nonzero mass.

If we set the mass equal to zero the given reasoning will lose their strength as in this case one can say nothing about the convergence of the integral (7) at the lower limit when $k^2 = 0$.

The author is greatly indebted to V.P.Gusynin for his attention to the work.

### REFERECES